# VISCOUS FLOW INSTABILITY OF INFLECTIONAL VELOCITY PROFILE

Hua-Shu Dou
Fluid Mechanics Division
National University of Singapore,
Department of Mechanical Engineering,
Singapore 119260
Email: mpedh@nus.edu.sg; huashudou@yahoo.com

**Abstract:** Rayleigh showed that inviscid flow is unstable if the velocity profile has an inflection point in parallel flows. However, whether viscous flows is unstable or not is still not proved so far when there is an inflection point in the velocity profile. Fluid viscosity has showed dual role to the flow instability. In this paper, it is demonstrated for the first time that viscous parallel flow with inflectional velocity profile is sufficient for flow instability.

## 1. INTRODUCTION

Rayleigh (1880)[1] first developed a general linear stability theory for inviscid plane-parallel shear flows, and showed that a necessary condition for instability is that the velocity profile has a point of inflection. Later, Tollmien (1935) succeeded in showing that this also constitutes a sufficient condition for the amplification of disturbances [2]. Fjϕrtoft (1950) gave a further necessary condition for inviscid instability, that there is a maximum of vorticity for instability [3-8]. Therefore, it is well known that inviscid flow with inflectional velocity profile is unstable, while inviscid flow with no inflectional velocity profile is stable. In contrast, it has not been theoretically shown so far whether or not viscous flow with inflectional velocity is unstable [9-10]. The associated analysis due to the effect of viscosity is more difficult than that in inviscid case [2-4]. Recently, Dou (2002) [11] proposed a new mechanism for flow instability and turbulent transition in parallel shear flows. In this paper, based on the previous work [11], it is demonstrated that parallel shear flows with inflectional velocity profile for two-dimensional and axisymmetrical viscous flows are sufficient for instability.

## 2. THEORY FOR FLOW INSTABILITY

The conservation of momentum combined with mass continuity for an incompressible Newtonian fluid (neglecting gravity force) can be expressed as,

$$\rho \frac{\partial \mathbf{V}}{\partial t} + \nabla\left( p + \frac{1}{2}\rho V^2 \right) = \mu \nabla^2 \mathbf{V} + \rho(\mathbf{V} \times \nabla \times \mathbf{V}), \qquad (1)$$

where $\rho$ is the fluid density, t the time, **V** the velocity vector, p the hydrodynamic pressure, $\mu$ the dynamic viscosity of the fluid.

Dou (2002) [11] demonstrated that the total energy gradient in the transverse direction of the main flow and the viscous friction in the streamwise direction dominate the instability and the flow transition. The energy gradient in the transverse direction tries to amplify a velocity disturbance, while the viscous friction loss in streamwise direction could resist and absorb this small disturbance. The transition of turbulence depends on the relative magnitude of the two roles of energy gradient disturbance and viscous friction damping. Based on above discussion, and referring to Eq.(1), a new dimensionless parameter, K, the ratio of the energy gradient in the transverse direction to the viscous force term in the streamwise direction, is defined as below (neglecting gravity force) [11],

$$K = \frac{\partial}{\partial n}\left(p + \frac{1}{2}\rho V^2\right) / \left(\mu \nabla^2 \mathbf{V}\right)_s, \qquad (2)$$

where n denotes the direction normal to the streamwise direction and s denotes the streamwise direction. The occurrence of instability depends upon the magnitude of this new dimensionless parameter K. The proposed principle can be used to both pressure and shear driven flows. But, it is assumed in this paper that there is no energy input (such as shear) to the system or energy output from the system. Actually, Fjϕrtoft (1950)'s criterion [5] is to exclude the shear-driven flow instability (for example, plane Couette flow) using Rayleigh's criterion. The parameter K in Eq.(2) is a field variable. For given flow geometry and fluid, when the maximum of K in the flow field is larger than a critical value Kc, instability would occur [11]. Dou (2002) demonstrated that the turbulence transition takes place at a critical value of Kc of about 385 for both plane Poiseuille flow and pipe Poiseuille flow, obtained excellent agreement with the experimental data. This result proved that the flow instability is resulted from the energy diffusion, but not the kind of eigenvalue instability of linear equations. Both Grossmann [12] and Trefethen et al.'s [13] commented that the nature of the onset-of-turbulence mechanism in parallel shear flows must be different from an eigenvalue instability of linear equations of small disturbance. From this result, it can be presumed that the transition of turbulence in other complicated shear flows would also depend on the maximum of K in the flow field.

## 3. TWO-DIMENSIONAL FLOW

For plane parallel flows (2D), $\partial p/\partial y=0$, the ratio of the energy gradient to the viscous force term, K, is,

$$K = \frac{\partial}{\partial y}\left(\frac{1}{2}\rho V^2\right) / \mu\left(\frac{\partial^2 u}{\partial y^2}\right). \qquad (3)$$

When there is no inflection point in the velocity distribution, K has a definite value in the flow field. Whether the flow is unstable depends on the magnitude of K value for given flow geometry and flow condition [11].

When there is an inflection point in the velocity distribution, $\partial^2 u/\partial y^2 = 0$, at the inflection point. If $\partial(0.5\rho V^2)/\partial y \neq 0$ at the inflection point, K will be infinite at this point and the flow is necessarily unstable. Therefore, viscous flows (2D) with inflectional velocity profile are unstable.

## 4. AXISYMMETRICAL FLOW

For axisymmetrical parallel flows, $\partial p/\partial r=0$, the ratio of the energy gradient to the viscous force term, K, is,

$$K = \frac{\partial}{\partial r}\left(\frac{1}{2}\rho V^2\right) / \mu\left(\frac{\partial^2 u_z}{\partial r^2} + \frac{1}{r}\frac{\partial u_z}{\partial r}\right) \qquad (4)$$

When there is no inflection point in the velocity distribution, $\mu(\partial^2 u_z/\partial r^2) < 0$, and $\mu(\partial^2 u_z/\partial r^2 + (1/r)\partial u_z/\partial r) < 0$ owing to flow loss. Thus, the magnitude of K has a definite value in the flow field. Whether the flow is unstable depends on the value of K for given flow geometry and flow condition.

When there is an inflection point in the axial velocity distribution, $\partial u_z/\partial r = 0$ at the inflection point; but $\mu(\partial^2 u_z/\partial r^2 + (1/r)\partial u_z/\partial r)$ may not be zero because there may be $(1/r)\partial u_z/\partial r \neq 0$ at the inflection point. At this case, we can prove that there is at least one point in the velocity profile at which $\mu(\partial^2 u_z/\partial r^2 + (1/r)\partial u_z/\partial r) = 0$.

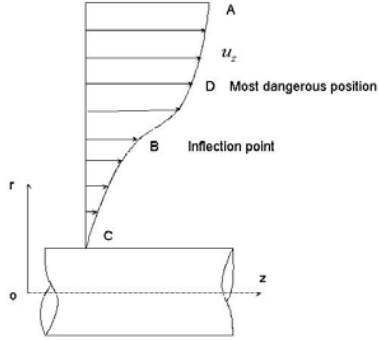 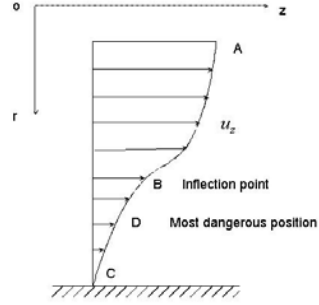

Fig. 1 Boundary layer on the convex surface for axisymmetrical flows when there is an inflection point in the velocity profile.

Fig. 2 Boundary layer on the concave surface for axisymmetrical flows when there is an inflection point in the velocity profile.

If the boundary layer is on the convex surface (Fig.1), $\mu(\partial^2 u_z/\partial r^2 + (1/r)\partial u_z/\partial r) > 0$ at the inflection point (B) owing to $(1/r)\partial u_z/\partial r > 0$. At a position which has an enough large distance from the wall, there must be $\mu(\partial^2 u_z/\partial r^2 + (1/r)\partial u_z/\partial r) < 0$ owing to flow loss (A). Therefore, there is at least one point in the velocity profile at which $\mu(\partial^2 u_z/\partial r^2 + (1/r)\partial u_z/\partial r) = 0$ (D). If $\partial(0.5\rho V^2)/\partial y \neq 0$ at this point, K will be infinite at this point and the flow is necessarily unstable. If the boundary layer is on the concave surface (Fig.2), there is $\mu(\partial^2 u_z/\partial r^2 + (1/r)\partial u_z/\partial r) < 0$ at the inflection point owing to $(1/r)\partial u_z/\partial r < 0$ (B). At the wall, there must be $\mu(\partial^2 u_z/\partial r^2 + (1/r)\partial u_z/\partial r) > 0$ owing to that there should be an inverse pressure gradient along the streamwise direction ∂p/∂x>0 (C). Therefore, there is at least one point in the velocity profile at which $\mu(\partial^2 u_z/\partial r^2 + (1/r)\partial u_z/\partial r) = 0$ (D). If $\partial(0.5\rho V^2)/\partial r \neq 0$ at this point, K will be infinite at this point and the flow is necessarily unstable. In summary, two-dimensional and axisymmetrical (along convex and concave surfaces) viscous flows with inflectional velocity profile are all unstable. The inviscid instability and viscous instability are compared in Table 1.

| Rayleigh (1880) | inviscid | no inflection | stable |
|---|---|---|---|
| | inviscid | inflection | unstable |
| Present theory | viscous | no inflection | $K_{max} \leq K_c$, stable |
| | viscous | no inflection | $K_{max} > K_c$, unstable |
| | viscous | Inflection | $K_{max} = \infty$, unstable |

Table 1 Criteria of stability of parallel flows for both two-dimensional and axisymmetrical flows. The value of $K_c$ depends on the flow geometry and fluid property. For both 2D and pipe Poiseuille flows, Kc=385[11].

## 5. DISCUSSION
The energy gradient in the transverse direction tries to amplify a small disturbance, while the energy loss due to friction in the streamwise direction plays a damping part to the disturbance. The parameter K represents the relative magnitude of disturbance amplification due to energy gradient and disturbance damping of viscous loss. When there is no inflection point in the velocity distribution, the amplification or decay of disturbance depends on the two roles above, i.e., the value of K. When there is an inflection point in the velocity distribution, the viscous term vanishes, and while the transversal energy gradient still exists at the position of inflection point. Thus, even a small disturbance must be amplified by the energy gradient at this location. Therefore, the flow will be unstable.

The theory can be used to explain a few flow problems. The flows around a cylinder displays a complicated sequence of flow patterns as the Reynolds number is increased [8,14]. For many years, it has been thought that the basic instability mechanism around a cylinder (Re exceeds 48.5) is likely to be inviscid because two inflection points exist in the wake [8,14] (from Rayleigh theorem). According to the present theory, this instability is of a viscous instability, rather than an inviscid instability. In terms of the present criterion, viscous flows with inflectional velocity profile is certainly unstable. We also notice that the Reynolds number at instability is very low (Re<50), the flow behaviour should not be inviscid, and viscous force must play an important role.

The instability mechanism can also get some hint from solid mechanics. As is well known, the damage of a metal component generally starts from some area such as manufacturing fault, crack, stress concentration, or fatigue position, etc. In fluid mechanics, the breaking down of a steady flow should also start from some most dangerous position first. For example, for the flow around an airfoil at a large attack angle, the flow instability first starts from the rear part on the suction side where the pressure gradient is large. For the flow around a cylinder described above, it is known that the flow instability begins first from the two inflection points near the rear stagnation point.

## 6.CONCLUSION
Sufficient large value of transversal energy gradient has the tendency to amplify a disturbance and may lead to flow instability. Streamwise energy loss tries to damp the disturbance and delay the instability. A dimensionless field variable K is used to characterize the relative magnitude of disturbance amplification and damping as well as the extent of instability. When the maximum of K in the flow field is larger than a critical value Kc, it is expected that the instability occurs. If there is an inflection point in the velocity profile, the value of maximum of K in the flow will be infinite and the flow becomes unstable. Following this principle, it is proved that viscous parallel flow with inflectional velocity profile is sufficient for flow instability for both two-dimensional and axisymmetrical flows.